# An $(m, k)$-firm Elevation Policy to Increase the Robustness of Time-Driven Schedules in 5G Time-Sensitive Networks

SIMON EGGER, University of Stuttgart, Germany
ROBIN LAIDIG, University of Stuttgart, Germany
HEIKO GEPPERT, University of Stuttgart, Germany
LUCAS HAUG, University of Stuttgart, Germany
JONA HERRMANN, University of Stuttgart, Germany
FRANK DÜRR, University of Stuttgart, Germany
CHRISTIAN BECKER, University of Stuttgart, Germany

Current standardization efforts are advancing the integration of 5G and Time-Sensitive Networking (TSN) to facilitate the deployment of safety-critical industrial applications that require real-time communication. However, there remains a fundamental disconnect between the probabilistic 5G delay characteristics and the often idealistic delay models used to synthesize 5G-TSN network configurations. For time-driven schedules in particular, any delay outlier unforeseen during schedule synthesis can jeopardize the robustness of their real-time guarantees. To address this challenge, we present the $(m, k)$-firm Elevation Policy to uphold a base level of weakly hard real-time guarantees during unstable network conditions that do not match the expected delay characteristics. It augments the primary time-driven schedule with a dynamic priority-driven scheme to elevate the priority of $m$ out of $k$ consecutive frames if they are delayed. Our evaluations demonstrate that weakly hard real-time guarantees are essential to uphold the quality of control within a networked control system. At the same time, only a small overhead is imposed when the primary schedule can provide stronger quality of service guarantees. Our $(m, k)$-firm Elevation Policy thereby yields a robust but light-weight fallback mechanism to serve applications with meaningful guarantees during unstable network conditions.

## 1 Introduction

Real-time communication is an important foundation to ensure a bounded network latency for control messages between sensors, controllers, and actuators in industrial applications [37]. Moreover, a wide range of modern industrial applications — like automated guided vehicles (AGVs) transporting materials and operating as part of a flexible production line — require a convergence of wired and wireless real-time networks to facilitate their seamless operation [5]. To ensure safety requirements, industrial applications have to meet certifiable safety integrity levels (SILs), which define a "tolerable risk", e.g., in form of an average frequency of dangerous failures per hour of operation [27]. This imposes stringent quality of service (QoS) requirements on the underlying network infrastructure that are often expressed in terms of per-stream latency and reliability.

Time-Sensitive Networking (TSN) introduces a standardized network technology stack as part of IEEE 802.1 that provides real-time guarantees in Ethernet networks. For periodic traffic, which typically has the most stringent latency and reliability requirements for closed-loop control tasks, TSN supports the Time-Aware Shaper (TAS) [17]. TAS provides a time-driven mechanism that enables network bridges to forward frames at precise times. The computation and configuration of this schedule is performed by a centralized network controller (CNC) [20], which has a global view of all network devices and stream specifications. Although prior fieldbus technologies like EtherCAT or SERCOS III provide similar time-division multiplexing capabilities, TSN offers a

Authors' Contact Information: Simon Egger, University of Stuttgart, Germany; Robin Laidig, University of Stuttgart, Germany; Heiko Geppert, University of Stuttgart, Germany; Lucas Haug, University of Stuttgart, Germany; Jona Herrmann, University of Stuttgart, Germany; Frank Dürr, University of Stuttgart, Germany; Christian Becker, University of Stuttgart, Germany, {simon.egger,robin.laidig,heiko.geppert,lucas.haug,jona.herrmann,frank.duerr,christian.becker}@ipvs.uni-stuttgart.de.

vendor-independent network deployment with novel avenues for interoperability. Notably, recent standardization efforts by 3GPP [2] aim to extend the real-time capabilities of TSN to networks with both wired and wireless network elements. They integrate the 5G system as a logical TSN bridge that supports fundamental TSN mechanisms (e.g., time synchronization and traffic shaping) and allows the CNC to configure the 5G system like any other wired TSN bridge.

A vast literature exists on traffic engineering in wired TSN networks: It ranges from solver-based approaches that yield optimal schedules [7, 35] to heuristic approaches that prioritize speed and scalability [9, 12]; a survey can be found in [38]. More recent approaches also introduce wireless-friendly scheduling techniques [10, 16] that utilize statistical knowledge of the 5G channel conditions. All have in common, however, that they assume well-defined *stable network conditions* that suit their (near-)deterministic delay models [7, 9, 12, 35], yield stationary 5G delays [16], or allow computing the $p$th-percentile of the 5G delays [10]. As realistic deployments are prone to abrupt delay anomalies that deviate from previous network conditions (e.g., due to 5G line-of-sight blockage or 5G handovers), we advocate for the deployment of a light-weight fallback strategy that serves applications with a meaningful base level of QoS if network conditions are unstable.

Within this context, we introduce a $(m, k)$-firm Elevation Policy to serve applications with weakly hard real-time guarantees, as defined by [3, 23]: Each stream specifies a $(m, k)$-firm latency requirement, stating that at least $m$ out of any $k$ consecutive frames must arrive at the listener before their deadline. Compared to percentage-based reliability metrics, e.g., 99 % of frames must meet their deadline [10], we argue that $(m, k)$-firm latency requirements can capture common survival time specifications more accurately. For instance, it captures the maximum time a system can operate correctly without receiving new messages [1] and the maximum allowed frame loss per unit of time [31]. In contrast, percentage-based reliability guarantees hold asymptotically and may eventually compensate for the loss of multiple consecutive frames over time.

The $(m, k)$-firm Elevation Policy realizes a dynamic priority-driven scheme that acts as a fallback to the primary time-driven schedule. It configures TSN bridges via so-called $\mu$-patterns to coordinate whether to elevate or discard an incoming frame when it arrives later than what is expected by the primary schedule. While elevated traffic is forwarded with the highest priority to minimize additional delays, it upholds the validity of the primary schedule by prolonging and deferring its scheduled transmission slots. We demonstrate that the $(m, k)$-firm Elevation Policy can be coupled with existing TSN schedulers to provide stronger QoS during stable network conditions and to verify these QoS guarantees remain satisfied after augmenting the schedule.

To summarize, the central contributions of our $(m, k)$-firm Elevation Policy are:

- A more meaningful QoS metric to capture common survival time specifications more accurately. Evaluations in our physical testbed show that weakly hard real-time guarantees can drastically improve the quality of control for a network-controlled inverted pendulum.
- A robust fallback mechanism to the primary schedule that upholds a base level of QoS during unstable network conditions. Our analysis in a 5G-TSN simulation environment shows it can sustain abrupt delay outliers that are close to exceeding the frame's deadline.
- An efficient schedule augmentation that remains complementary to existing TSN schedulers. We demonstrate that it is compatible with conventional and wireless-friendly schedulers, while imposing only a small latency overhead during stable network conditions.

The remainder of this paper is structured as follows: Section 2 discusses related work. Section 3 provides further background on TSN and its integration with 5G. Section 4 defines our system model and summarizes the problem statement. Section 5 introduces the $(m, k)$-firm Elevation Policy and how to facilitate its correct deployment. Section 6 evaluates and analyzes its performance when deployed in our physical testbed and a simulation environment. Section 7 reflects on open limitations and discusses practical aspects. Finally, Section 8 concludes this work.

## 2 Related Work

We structure the related work discussion in accordance with the central contributions of this work.

*Weakly Hard Real-Time Guarantees.* Both within and outside TSN research, scheduling research often considers percentage-based reliability metrics [10, 29] or weighted-sum objectives to penalize deadline violations [40] for networks with stochastic delay characteristics. The downside of these metrics is that they permit large bursts of frame losses (if the loss ratio is averaged out over time) that can significantly impair the applications' performance. In contrast, this work considers weakly hard real-time guarantees [3] for streams with $(m,k)$-firm deadlines [23]. By requiring that at least $m$ out of $k$ consecutive frames meet their deadlines, this model prohibits large bursts of frame losses and captures survival time requirements of industrial applications more accurately [1, 31]. Moreover, it was shown that (compared to percentage-based reliability metrics) weakly hard real-time constraints result in superior stability guarantees for networked control systems [4].

*Robust QoS Guarantees in Converged Wired and Wireless Time-Sensitive Networks.* Traffic engineering research in wired TSN often assumes deterministic delay characteristics to synthesize schedules with hard real-time guarantees, i.e., where every single frame meets its deadline [7, 9, 26, 35]. Applying these techniques to networks with wireless links results in poor efficiency, as they are not designed to tolerate large delay variations [10]. The authors of [10] thus propose the usage of 5G delay histograms to provide per-stream latency bounds with asymptotic reliability guarantees. However, this approach tightly couples the ultra-high reliability guarantees (e.g., 99.999 %) with the accuracy of the histograms and is therefore prone to abrupt 5G delay disturbances (e.g., in cases of 5G line-of-sight blockage or 5G handovers). In contrast, our $(m,k)$-firm Elevation Policy is designed to work as a fallback mechanism to the primary schedule — i.e., which can be synthesized by any of the above schedulers — to improve their robustness against delay outliers.

A second line of TSN research focuses on improving reliability under different fault models (e.g., sporadic frame loss or permanent link failures). These reliability mechanisms can be categorized into spacial or temporal redundancy strategies: Spacial redundancy strategies carefully select disjoint routes for frame replicas [11, 15, 45] and remain largely applicable in our setting, e.g., as studied in [42]. However, even with the necessary infrastructure (i.e., one 5G UE and 5G UPF per redundant wireless path), spacial redundancy cannot provide weakly hard real-time guarantees on its own without support from the underlying traffic scheduler. Out of the temporal redundancy strategies [8, 13, 44], we argue that the work of [44] is most closely related to our approach. While the authors account for sporadic delays in the release time of frames, this work designs a more general fallback mechanism to detect and mitigate delay anomalies in any network segment. Even in the special case of sporadic release times, our evaluations show that the $(m,k)$-firm Elevation Policy can improve schedulability and reduce the resource overhead on top of the primary schedule.

*Complementary Scheduler Design.* We argue that the complementary design of our $(m,k)$-firm Elevation Policy is fundamental to realize dependable network updates in 5G-TSN. A related concept of *per-packet consistent network updates* exists already for Software-Defined Networks (SDN) to guarantee certain properties (e.g., no forwarding loops or black holes) are satisfied before, during, and after network updates [32]. 5G-TSN networks benefit from similar consistency guarantees in terms of QoS when the network needs to be reconfigured, e.g., if the 5G delay characteristics change permanently due to increased traffic load. Still, to the best of our knowledge, the closes attempt in TSN literature remains limited to the scheduling *flexibility* to accommodate additional

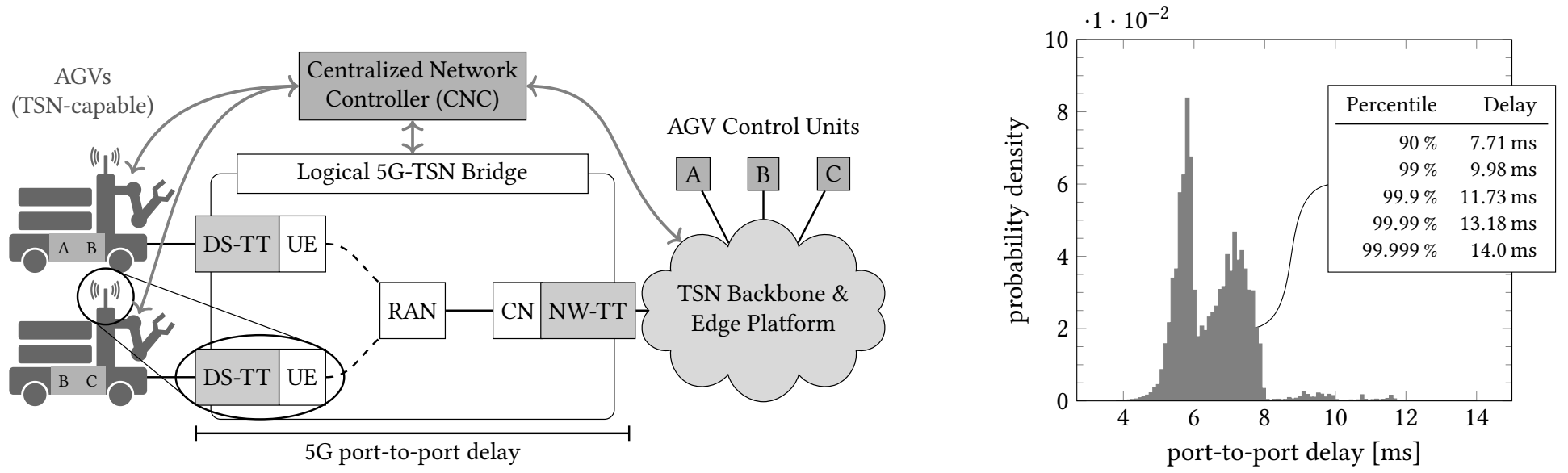

(a) The 5G system as a logical 5G-TSN bridge

(b) 5G uplink delays, measured by [25]

Fig. 1. Converged wired and wireless networks enable modern industrial applications (a), but providing formal QoS guarantees under the significant 5G delay variations (b) remains a challenging task.

streams in the future [14]. In contrast, our $(m, k)$-firm Elevation Policy can uphold a base level of weakly hard real-time guarantees until the new network schedule is deployed and operational.

## 3 Background

The integration of 5G and TSN enables time-sensitive industrial applications that rely on flexible deployment, device mobility, and real-time communication. This section provides background on 5G-TSN and introduces cooperative automated guided vehicles as a representative use case that requires weakly hard real-time guarantees.

### 3.1 Use Case: Cooperative Automated Guided Vehicles (AGVs)

In a mobile production line, a fleet of AGVs can both participate in manufacturing (e.g., by being docked at the production line) and transport materials to and from the warehouse. The underlying network and compute infrastructure is illustrated in Fig. 1a: Each AGV is equipped with multiple sensors and actuators — which may also be connected via an internal wired TSN network — and a 5G user equipment (UE) module that connects wirelessly to the 5G radio access network (RAN). The 5G core network (CN) connects to the wired TSN backbone to forward traffic to and from the on-site edge compute platform. Different AGV control units can be hosted at the edge, e.g., to control/monitor the movement of individual AGVs (e.g., unit A and C) and to coordinate the operation of the entire AGV fleet (e.g., unit B, which is deployed on all AGVs).

Each networked control loop introduces network requirements that are specified as one or multiple TSN streams. For this work, we focus on time-triggered streams as they are commonly attributed with the most stringent requirements for safety-critical applications (e.g., movement coordination and safety stops). A time-triggered streams is characterized by (i) a fixed transmission period, (ii) a maximum frame size, and (iii) QoS requirements. While QoS for wired networks is mainly expressed in terms of latency, use case descriptions for converged wired and wireless industrial networks often depend on a *survival time*, i.e., the maximum time the system can endure communication faults [1, 31]. This notion of survival time can be expressed by $(m, k)$-firm requirements: For instance, while some applications allow zero faults (i.e., $m = 1$, $k = 1$), other applications can tolerate $n$ consecutive faults (i.e., $m = 1$, $k = n + 1$).

### 3.2 Time-Sensitive Networking (TSN)

TSN provides a suite of IEEE standards that specify protocols and mechanisms for time synchronization [21], frame replication for improved reliability [19], and traffic shaping/policing to provision real-time guarantees for time-critical traffic [22]. Ethernet frames associated with a TSN stream use an IEEE 802.1Q header field that specifies a VLAN ID and priority code point (PCP) value. These specifications allow TSN bridges to map incoming frames to their stream specification and to determine the priority with which they should be forwarded.

For this work, we assume that the clocks of all network devices are synchronized (max. 1 µs clock skew [21]) and focus on mainly two TSN mechanisms: the Time Aware Shaper (TAS) [17] and Per-Stream Filtering and Policing (PSFP) [18]. Their functionality is shown in Fig. 2 for a simplified case with a single ingress and egress port. When a frame arrives at the TSN bridge, PSFP first verifies if the frame adheres to the configured TSN schedule. For (scheduled) periodic traffic, PSFP is typically configured to enqueue the frame (e.g., identified by its header fields) if and only if it arrives within an expected time interval; otherwise, the frame is discarded. After passing PSFP, the TSN bridge enqueues the frame as follows: The frame's destination MAC address determines the correct egress port (in Fig. 2, there is only one) and the frame's 3-bit PCP value determines the correct FIFO queue. TAS then uses a so-called gate control list (GCL) to govern when the gate assigned to an egress queue is open or closed. A frame is selected for transmission if (1) it is at the head of the FIFO queue, (2) the gate assigned to the queue is open, and (3) there exists no frame in a higher-priority queue that is eligible for transmission at the same time.

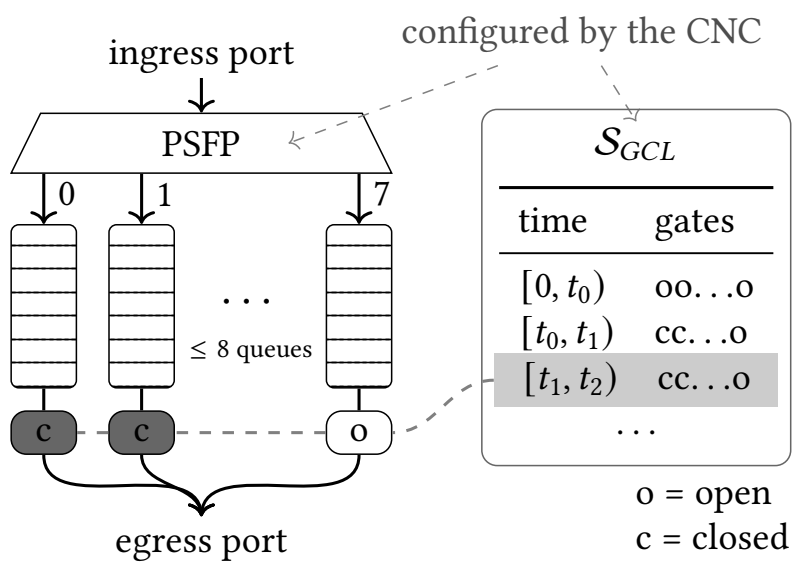

Fig. 2. Port-to-port model of a TSN bridge with PSFP and TAS.

### 3.3 Integration of 5G and TSN

Recent standardization efforts by 3GPP enable the architectural integration of the 5G system as a logical 5G-TSN bridge [2]. As depicted in Fig. 1a, this integration introduces an important separation between the physical and the logical network entities: While each AGV is fitted with a 5G UE, the logical 5G-TSN bridge groups together the 5G UEs, RAN, and CN into a single logical entity to interface with the TSN controller. This abstraction allows the 5G system to hide internal state information from the TSN control plane (e.g., 5G resource allocation and session management), while enabling the same user plane functionality on the ingress and egress side as a wired TSN bridge. In particular, the ingress and egress sides are equipped with so-called device-side (DS-TT) and network-side TSN translators (NW-TT) to support the TSN mechanisms from above.

Still, there remains a fundamental discrepancy in the delay characteristics of wired TSN bridges and logical 5G-TSN bridges. This is shown most prominently by the bridges' *port-to-port delays*, which capture the time it takes for a frame to traverse the bridge from ingress to egress port (excluding any queuing and cross-traffic delay). For wired TSN bridges, these delays are often in the range of microseconds, with variations in the range of a few hundred nanoseconds. For the 5G system, measurements [25] show that both the absolute delays and the delay variations are three orders of magnitude larger (see Fig. 1b). Moreover, the 5G delay characteristics may change over time (e.g., due to changing traffic load) or even abruptly (e.g., due to blockage or handovers).

## 4 System Model and Problem Statement

This section starts by introducing general definitions and our modeling assumptions that are used throughout the paper. Thereafter, we discuss why existing time-driven scheduling techniques are not sufficient to serve $(m, k)$-firm latency guarantees in 5G-TSN.

## 4.1 General Definitions and Notation

*Network Graph.* We model the network as a directed graph $G = (V, E)$. Vertices $u \in V$ denote network devices that are visible to the CNC, namely wired TSN bridges, DS-TTs, NW-TTs, and TSN end devices. Edges $(u, v) \in E$, for $u, v \in V$, represent full-duplex Ethernet links or 5G links. We assume that the CNC has a complete view of $G$, i.e., devices forward their aggregated LLDP information to the CNC, and that $G$ does not change at runtime. Note that 5G handovers are not visible in terms of a changing network topology, as both the DS-TT and NW-TT remain the same. Instead, 5G handovers result in higher 5G delays for a limited timespan.

*Traffic Specification.* We consider a fixed set of time-triggered streams $\mathcal{F}$. Each stream $F \in \mathcal{F}$ specifies its route $F.route = (v_F^1, v_F^2, \dots, v_F^{n_F})$, its priority in form of a PCP value $F.pcp$, its transmission period $F.period$, its offset within the period $F.phase$, and its maximum frame size $F.size$ in bits. The end devices in $F.route$ are referred to as talker $v_F^1$ and listener $v_F^{n_F}$. The $i$th frame of $F$, denoted by $f^{(i)} \in F$, is released by the application at time

$$f^{(i)}.release := F.phase + (i-1) \times F.period$$

and meets the latency requirement $F.lat$ (with $F.lat \leq F.period$) if and only if

$$f^{(i)} \text{ arrives at the listener before } f^{(i)}.release + F.lat. \tag{1}$$

We only use the verbose notation $f^{(i)}$ when the index $i$ matters. Moreover, we may simply write $f.period$ instead of $F.period$ (and so forth) to avoid explicitly referring to the stream $F$.

*TSN Schedules.* A TSN schedule $C = (\mathcal{R}, \mathcal{S}_{GCL})$ specifies the expected arrival intervals $\mathcal{R}$ for PSFP and the gate control lists $\mathcal{S}_{GCL}$ for TAS. As detailed in Section 3.2, PSFP is configured for each TSN bridge, whereas TAS is configured for each egress port. When referring to a specific entry, we therefore often write $\mathcal{R}(u)$ to denote the PSFP configuration at bridge $u \in V$ and $\mathcal{S}_{GCL}(u, v)$ to denote the GCL configuration at the port $(u, v) \in E$. Finally, note that both $\mathcal{R}$ and $\mathcal{S}_{GCL}$ are finite as the schedule repeats after each *hypercycle* $H = \text{lcm}_{F \in \mathcal{F}}(F.period)$.

## 4.2 Problem Statement

Due to its time-driven nature, every TSN schedule is synthesized based on certain delay assumptions. These assumptions range from being deterministic, e.g.,

(A1) every frame $f$ is available for transmission exactly at the time $f.release$ [7, 9, 10, 26, 35], or
(A2) wired per-link transmission delays of $f$ can be expressed by fixed time slots [7, 9, 26, 35],

to being more probabilistic to account for bounded delay variations, e.g.,

(A3) the 5G delays of $f$ are bounded by $[d^{min}, d^{max}]$ with a probability $\alpha$ [10].

The validity of assumptions (A1) and (A3) in particular are difficult to justify in more complex networks. For instance, there can be unexpected compute delays or time synchronization errors that violate (A1) or abrupt 5G delay outliers (e.g., due to blockage or handovers) that violate (A3). Moreover, (A3) is subjected to long-term changes (e.g., due to changing traffic load) that can prompt the CNC to update the deployed TSN schedule, which involves

(1) recomputing the confidence intervals for the 5G delay percentiles,
(2) recomputing the TSN schedule to satisfy the QoS requirements of each stream, and
(3) deploying the new GCL and PSFP configurations on all TSN devices (e.g., via NETCONF).

These steps take considerable time (possibly seconds) and can result in network downtime.

For this work, we consider a primary TSN schedule $C$ that serves each stream $F \in \mathcal{F}$ with certain QoS guarantees during stable network conditions, i.e., epochs where the delay assumptions of $C$ hold true for every frame $f \in F$. We design a fallback policy to uphold a base level of weakly hard real-time guarantees if network conditions become unstable. Each stream $F \in \mathcal{F}$ is allowed to specify an $(m, k)$-firm latency requirement, stating that at least $m$ out of any $k$ consecutive frames have to satisfy the latency bound of Eq. (1). Still, it is clear that no real-time guarantees are possible for *arbitrary* delay outliers, e.g., in case the outlier itself already exceeds the frame's deadline $f.release + f.lat$. The fallback policy must therefore specify bounds on the endurable delay outliers to provide dependable guarantees and to shield the QoS guarantees of other time-sensitive traffic. Finally, we want to evaluate the imposed resource overhead of the fallback policy during stable network conditions where the primary TSN schedule can provide stronger QoS guarantees.

## 5 Network Architecture and Scheduler Design

This section proposes the $(m, k)$-firm Elevation Policy to achieve the above design goals. Our realization of this policy is guided by the answers to the following two design questions:

*1) How to handle delayed frames during unstable network conditions?* While scheduled traffic can be forwarded with minimal network latency in TSN, its correct operation can only be assured as long as every frame is forwarded during its intended transmission slots at each port. Our $(m, k)$-firm Elevation Policy therefore augments the primary schedule $C$ with a dynamic priority-driven scheme to elevate the priority of a delayed frame $f \in F$ if and only if $f$ endangers the $(m, k)$-firm requirement of $F$; otherwise, $f$ is discarded via PSFP to shield the QoS of other scheduled traffic.

*2) How does it affect the streams' QoS during stable network conditions?* Ideally, the same stable QoS requests of the primary TSN schedule $C$ can be maintained. However, under the $(m, k)$-firm Elevation Policy, $C$ must also account for further blocking delays that are caused by late frames with elevated priority. Section 5.2 introduces a simple augmentation of $C$ that prolongs and defers its scheduled transmission slots accordingly, while staying close to the initial schedule. We define this procedure as a one-shot routine for candidate solutions, which can serve as feedback to the scheduler in case a QoS violation is detected. For instance, it can prompt rescheduling to try finding an improved solution or a graceful degradation in the streams' QoS (during stable conditions).

### 5.1 Network Architecture to Support the $(m, k)$-firm Elevation Policy

*5.1.1 The $(m, k)$-firm Elevation Policy.* On first sight, it appears natural to reactively elevate the priority of the $i$th frame $f^{(i)} \in F$ whenever its loss would violate the $(m, k)$-firm requirement of $F$. However, applying this reactive policy with a restricted local view of the network can lead to unintended side effects when multiple network segments cause unexpected delay outliers:

Consider a stream $F$ with a $(1, 2)$-firm latency requirement and the route shown in Fig. 3a. To improve robustness against delay outliers from sporadic release times and 5G channel degradations, the reactive policy is employed at $B_1$ and $B_{NW}$. Now, let $f^{(i-1)}, f^{(i)} \in F$ be two consecutive frames and assume the following scenario:

- $f^{(i-1)}$ passes $B_1$ normally but is discarded by $B_{NW}$ due to an unexpected 5G delay.
- $f^{(i)}$ is discarded by $B_1$ due to an unexpected delay in its release time.

In particular, $B_1$ is unaware of $f^{(i-1)}$ being lost and wrongfully assumes that $f^{(i)}$ can be safely discarded. This results in a violation of the $(1, 2)$-firm latency requirement of $F$.

To circumvent this issue, we design the $(m, k)$-firm Elevation Policy to coordinate the priority elevation at $B_1$ and $B_{NW}$. In detail, we consider a fixed *$\mu$-pattern* that specifies which frames are eligible for priority elevation. The $\mu$-pattern for an $(m, k)$-firm latency requirement is defined as a $k$-bit word $\mu = \mu_0 \cdots \mu_{k-1}$ over the alphabet $\mu_i \in \{0, 1\}$ with $\mu_0 + \cdots + \mu_{k-1} \geq m$. When the $i$th frame $f^{(i)} \in F$ is delayed, its priority is allowed to be elevated if and only $\mu_{i \bmod k} = 1$.

For the previous example, configuring $B_1$ and $B_{NW}$ under $\mu = 01$ results in:

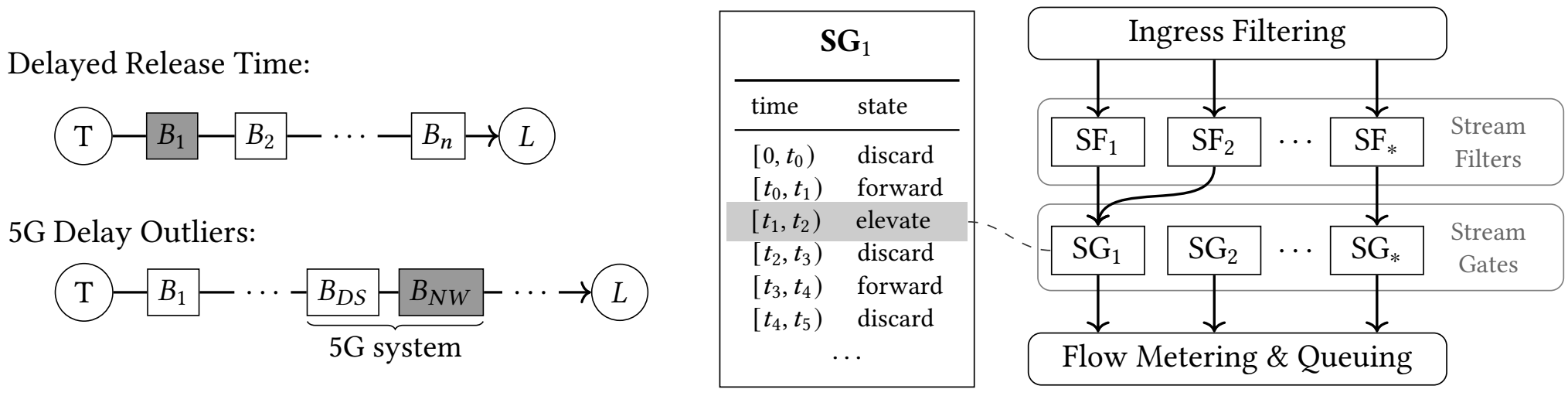

(a) Example elevation points (gray)

(b) Extended PSFP capabilities of TSN bridges

Fig. 3. The $(m, k)$-firm Elevation Policy is deployed on bridges that follow network segments prone to unexpected delays (a). These bridges require a simple mechanism to elevate the priority of delayed frames (b).

- Case ($i$ mod $2 = 0$): $B_{NW}$ elevates $f^{(i-1)}$ due to $\mu_{i-1 \text{ mod } 2} = \mu_1 = 1$ instead of discarding it.
- Case ($i$ mod $2 = 1$): $B_1$ elevates $f^{(i)}$ due to $\mu_{i \text{ mod } 2} = \mu_1 = 1$ instead of discarding it.

Neither case results in the same $(1, 2)$-firm latency violation from before.

*5.1.2 Extended TSN Bridge Capabilities.* To realize the $(m, k)$-firm Elevation Policy, we build on the mechanisms provided by Per-Stream Filtering and Policing (PSFP) [18] to stay as close as possible to the current TSN standards. Standard PSFP operates on the ingress side of TSN bridges and can be configured to discard frames that arrive outside an expected time interval. As shown in Fig. 3b, this behavior is implemented by TSN bridges via so-called *stream filters* and *stream gates*: Stream filters match incoming frames according to their header fields (e.g., src/dst MAC address and VLAN tag), whereas the stream gate (referenced by the first matching stream filter) determines if PSFP forwards or discards the frame. Moreover, PSFP supports bounding the maximum number of bytes that are allowed to pass during the *forward* state, after which it switches automatically to the *discard* state.

We extend PSFP by introducing an additional state for the stream gates, which we call *elevate*. It instructs PSFP to overwrite the PCP field of an incoming frame to the highest priority (i.e., $111_2 = 7$). This priority change stays visible for all subsequent bridges in the frame's route. We note that the CNC has full control over the PSFP configuration and can therefore specify exact intervals during which incoming frames are elevated, discarded, or forwarded normally. To illustrate, consider a stream $F$ that has an $(1, 2)$-firm latency requirement and matches the stream filter $SF_1$ in Fig. 3b. For the first two frames $f^{(0)}, f^{(1)} \in F$, the configuration of $SG_1$ yields the following semantics:

- If $f^{(0)}$ and $f^{(1)}$ arrive within their expected arrival interval $[t_0, t_1)$ and $[t_3, t_4)$, the primary TSN schedule is expected to serve $F$ without changing $F.pcp$.
- Otherwise, $SG_1$ switches to the fallback policy of serving the $(1, 2)$-firm latency requirement of $F$ by elevating the priority of $f^{(0)}$ when it arrives within the interval $[t_1, t_2)$.

For completeness, we note that no guarantees can be made if $f^{(0)}$ arrives outside the interval $[t_0, t_2)$. At worst, it may be discarded by PSFP or may even masquerade as $f^{(1)}$. Still, there exist techniques to isolate these effects from spreading to other streams (as demonstrated later in Section 6.3).

## 5.2 TSN Configurations for the $(m, k)$-firm Elevation Policy

Next, we show how to compute a suitable configuration for the $(m, k)$-firm Elevation Policy to achieve the above semantics. Throughout this section, we use Fig. 4 to illustrate the procedure in a simple network with four streams $F_1 - F_4$: Each stream $F_i$ originates from the talker $T_i$ and is forwarded to a common listener $L$. We assume that the wired streams $F_3$ and $F_4$ do not require $(m, k)$-firm latency guarantees, i.e., the talkers $T_3$ and $T_4$ guarantee their timely frame releases and

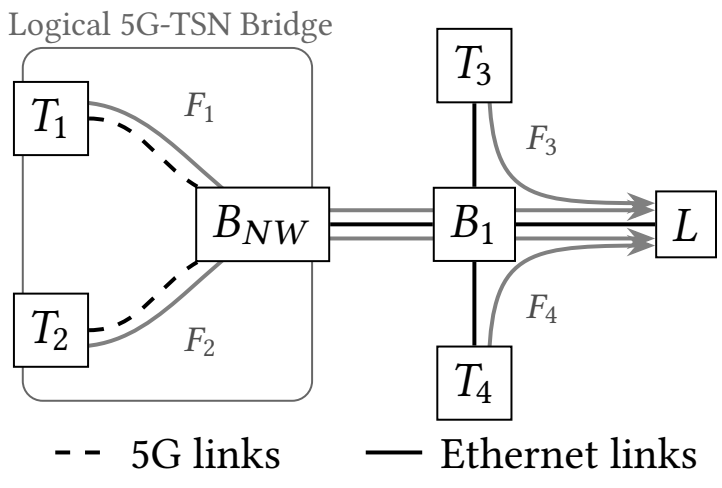

| Stream | PCP | Delay Variations | $(m, k)$-firm Req. |
|---|---|---|---|
| $F_1$ | 6 | 5G system | $(1, 3)$ |
| $F_2$ | 4 | 5G system | none* |
| $F_3$ | 5 | negligible | not needed* |
| $F_4$ | 6 | negligible | not needed* |

* corresponds to a $(0, 1)$-firm requirement

Fig. 4. Example network topology (left) with the stream specifications of $F_1$–$F_4$ (right).

there are no unexpected delays on their wired routes. In contrast, the streams $F_1$ and $F_2$ traverse a 5G system that is prone to abrupt 5G channel degradations (e.g., caused blockage or handover scenarios). With the primary TSN schedule $C$ serving $F_1$–$F_4$ during stable network conditions, it remains to augment $C$ to uphold the $(1, 3)$-firm latency requirement of $F_1$ throughout unstable network conditions and to quantify the maximum tolerated delay outliers.

We continue this section as follows: First, we introduce a token bucket specification to determine how much elevated traffic $C$ has to account for at each port. Thereafter, the token bucket specifications are used to *prolong* and *defer* the scheduled transmission slots in $C$ to account for the worst-case delays that can be induced by late frames with elevated priority.

*5.2.1 Modelling Elevated Traffic as Per-Link Token Buckets.* Let $\mathcal{F}_{(u,v)} \subseteq \mathcal{F}$ denote the streams that traverse the link $(u, v)$. We start by defining $N_F([t_1, t_2])$ to count how many frames of $F \in \mathcal{F}$ can be elevated within a time interval $[t_1, t_2]$: Let $\mu = \mu_0 \cdots \mu_{k-1}$ denote the $\mu$-pattern of $F$ and define

$$N_F([t_1, t_2]) = \mu_{l \bmod k} + \cdots + \mu_{h \bmod k}, \quad \text{with } l = \left\lfloor \frac{t_1 - F.lat}{F.period} \right\rfloor + 1 \text{ and } h = \left\lfloor \frac{t_2}{F.period} \right\rfloor,$$

where $l$ and $h$ determine the lowest and highest frame indices that can be elevated during $[t_1, t_2]$. The token bucket $TB(u, v)$ is defined to bound the arrival rate of elevated traffic at $(u, v)$ via

$$\text{Bucket Size:} \quad b(u, v) = \max \Big\{ \sum_{F \in \mathcal{F}_{(u,v)}} F.size \times N_F([t, t]) \mid 0 \leq t < H \Big\},$$

$$\text{Token Rate:} \quad r(u, v) = \max \left\{ \frac{\sum_{F \in \mathcal{F}_{(u,v)}} F.size \times N_F([t_1, t_2]) - b(u, v)}{t_2 - t_1} \mid 0 \leq t_1 < t_2 < 2H \right\},$$

Here, the bucket size $b$ is chosen conservatively to bound any simultaneous arrival of elevated frames from different streams. Complementary, the token rate $r$ ensures that there are always enough tokens left when an elevated frame arrives; note that $H \leq t_2 < 2H$ must be considered to account for overflows between two consecutive hypercycles $H$.

For example, $TB(B_1, L)$ only has to consider for $F_1$ in Fig. 4. All other streams $F \in \{F_2, F_3, F_4\}$ have a $\mu$-pattern of $\mu = 0$, which yields $N_F([t_1, t_2]) = 0$ for any interval $[t_1, t_2]$. Hence,

$$b(B_1, L) = F_1.size \quad \text{and} \quad r(B_1, L) = F_1.size/(3F_1.period - F_1.lat),$$

for any pattern $\mu \in \{001, 010, 100\}$ for $F_1$. The bucket size accounts for a single elevated frame of $F_1$, whereas the bucket needs to be refilled after the minimal inter-arrival time $3F_1.period - F_1.lat$.

*5.2.2 Augmenting TSN Schedules.* The primary TSN schedule $C$ specifies an ordering of frame transmissions at every egress port in the network. In the following, we augment $C$ to support the

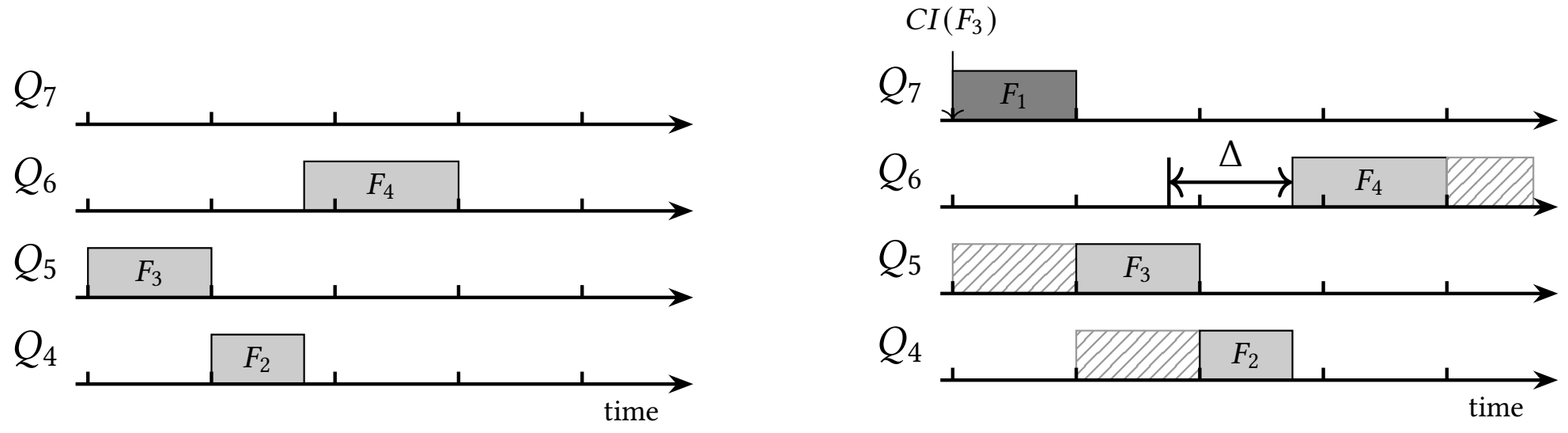

(a) Initial $\mathcal{S}_{GCL}(B_1, L)$ (b) Augmented $\tilde{\mathcal{S}}_{GCL}(B_1, L)$

Fig. 5. Illustration of the augmentation procedure for the bottleneck link $(B_1, L)$ from Fig. 4. The prolonged transmission slots are shown by shaded boxes, during which the gate is open but the transmission is either blocked (for $F_3$ and $F_2$) or has already completed (for $F_4$). The deferred transmission of $F_4$ is marked by $\Delta$.

$(m, k)$-firm Elevation Policy at a single port $(u, v) \in E$:[1] Let $(f_i, [o_i, c_i])_{i=1}^{N}$ denote the sequence of $N$ scheduled transmissions under $C$, i.e., the frame $f_i$ is scheduled for transmission during the interval $[o_i, c_i]$. W.l.o.g., we assume that the sequence is ordered by $o_1 \leq o_2 \leq \ldots \leq o_N$. We iteratively augmented the transmission intervals, denoted by $[\tilde{o}_i, \tilde{c}_i]$, through prolongation (i.e., $c_i - o_i \leq \tilde{c}_i - \tilde{o}_i$) and deferment (i.e., $o_i \leq \tilde{o}_i$).

To illustrate the steps of the procedure, we utilize the initial and augmented TSN schedule shown in Fig. 5: It shows an excerpt of $\mathcal{S}_{GCL}$ at the bottleneck link $(B_1, L)$, where the depicted time window captures the back-to-back transmission slots allocated for the streams $F_3$, $F_2$, and $F_4$. According to the stream's priority, the transmission slot refers to the interval during which the gate of the respective egress queue is opened. For instance, the gray box labeled by $F_3$ shows the interval where the gate of queue $Q_5$ is open (due to $F_3.pcp = 5$).

*Iteration 0 (Initialization):* The intervals are initialized with $[\tilde{o}_i, \tilde{c}_i] \leftarrow [o_i, c_i]$ (i.e., for $i = 1, \ldots, N$). The augmented TSN schedule is denoted by $\tilde{C} = (\tilde{\mathcal{S}}_{GCL}, \tilde{\mathcal{R}})$ (initially left empty) and populated step by step. Moreover, any variable with zero index (i.e., $\tilde{o}_0$, $\tilde{c}_0$, $f_0.size$, and $\Delta_0$) should be considered as a helper variable that is set to 0 to reduce the number of case distinctions.

*Iteration i (Prolongation):* There are two cases where elevated traffic can delay the frame $f_i$: First, elevated traffic can delay $f_i$ *directly* by arriving simultaneously to the gate opening time $\tilde{o}_i$. At worst, $f_i$ can be delayed by the maximum burst time of $TB(u, v)$, i.e.,

$$\theta_i^1 \leftarrow \tilde{o}_i + \frac{b(u,v)}{(u,v).bitrate - r(u,v)}.$$

For example, Fig. 5 shows this prolongation for $F_3$. The critical instance occurs at time $CI(F_3)$ when elevated traffic arrives at the time the gate is opened for $F_3$. In this simple scenario, the maximum burst time is approximately the serialization delay $F_1.size/(B_1, L).bitrate$.

Second, the transmission of $f_i$ can be delayed *indirectly* through the previous transmission of $f_{i-1}$ with overlapping intervals $[\tilde{o}_{i-1}, \tilde{c}_{i-1}] \cap [\tilde{o}_i, \tilde{c}_i] \neq \emptyset$. At worst, $f_{i-1}$ completes its transmission at time $\tilde{c}_{i-1}$ and there are are new tokens in $TB(u, v)$ that accumulated during the transmission of $f_{i-1}$. In this case, the transmission start $f_i$ is delayed until

$$\theta_i^2 \leftarrow \tilde{c}_{i-1} + \frac{(f_{i-1}.size/(u,v).bitrate) \times r(u,v)}{(u,v).bitrate - r(u,v)}.$$

[1]The multi-hop setting is analogous, which we provide in Appendix B for completeness.

For example, Fig. 5 shows this case for $F_2$. The prolongation of $F_3$'s transmission slot causes an overlap with the slot intended for $F_2$. As $F_3$ has precedence, the transmission start of $F_2$ can be delayed until the completion of $F_3$. While the second term of $\theta_i^2$ is negligible in this case, it accounts for accumulated tokens during the transmission of $F_3$, which grows for longer transmission chains.

Hence, to account for both cases, we prolong the transmission interval of $f_i$ until

$$\tilde{c}_i \leftarrow \theta_i + \frac{f_i.size}{(u,v).bitrate}, \quad \text{with } \theta_i \leftarrow \max\{\theta_i^1, \theta_i^2\}.$$

*Iteration i (Deferment):* A special case arises if the priority of $f_{i+1}$ is higher than that of $f_i$. Then, the prolongation of $[\tilde{o}_i, \tilde{c}_i]$ can result in the case where both $f_i$ and $f_{i+1}$ are simultaneously available for transmission. Thereupon, the transmission selection in TSN would choose $f_{i+1}$ for the subsequent transmission. We retain the initial transmission ordering of $C$ by updating

$$\Delta_i \leftarrow \Delta_{i-1} + \max\{0, \tilde{c}_i - \tilde{o}_{i+1}\}$$

and deferring the transmission of $f_{i+1}$ by $\tilde{o}_{i+1} \leftarrow \tilde{o}_{i+1} + \Delta_i$.

For example, Fig. 5 shows the deferment of $F_4$. This ensures that frames of $F_4$ do not interfere with earlier lower-priority transmission slots (i.e., of $F_3$ and $F_2$). Thereby, the above prolongation of $F_3$ and $F_2$ only has to account for $F_1$ and the initial transmission ordering is preserved.

*Iteration i (TSN Configuration Entry):* The following GCL and PSFP entry are added for $f_i$: $\tilde{\mathcal{S}}_{GCL}(u,v)$ opens the gate that is associated with the priority $f_i.pcp$ during the interval $[\tilde{o}_i, \tilde{c}_i]$. Similarly, $\tilde{\mathcal{R}}(v)$ adds an entry to forward $f_i$ if it arrives during the half-open interval

$$\big[\tilde{o}_i + d_{min}((u,v), f_i),\ \theta_i + d_{max}((u,v), f_i)\big). \tag{2}$$

Compared to the serialization delay $f_i.size/(u,v).bitrate$, we denote $[d_{min}, d_{max}]$ to account for the total delay between $f_i$'s transmission start at $u$ and its arrival at $v$ during stable network conditions.[2] In contrast, if $f_i$ is eligible for priority elevation under its configured $\mu$-pattern, we configure $\tilde{\mathcal{R}}(v)$ to elevate $f_i$ if it is delayed and arrives during the half-open interval

$$\big[\theta_i + d_{max}((u,v), f_i),\ f_i.release + f_i.lat\big). \tag{3}$$

In other words, the priority of $f_i$ is elevated if it arrives later than (2) but still has a chance of arriving at the listener before its deadline $f_i.release + f_i.lat$. However, if it arrives at $v$ even later than (3), PSFP discards $f_i$ to ensure the validity of the token bucket specification from Section 5.2.1.

For example, consider a frame $f_1 \in F_1$ that can be elevated under the $\mu$-pattern of $F_1$. The above PSFP configuration (2)–(3) is deployed for every bridge along $F_1.route = (T_1, B_{NW}, B_1, L)$: $f_1$ is thereby elevated at $B_{NW}$ if its 5G delay exceeds $d_{max}((T_1, B_{NW}), f_1)$. At the same time, it will be discarded by $B_{NW}$ or $B_1$ if it arrives later than $f_1.release + f_1.lat$ to ensure its elevation does not violate the token bucket specification of Section 5.2.1. Finally, we note that $B_1$ does not require PSFP if (3) is closed earlier at $B_{NW}$ (i.e., at time $f_1.release + f_1.lat - \epsilon$), accounting for $f_1$'s maximum transmission delay at $(B_{NW}, B_1)$. While $\epsilon$ can be computed with network calculus tools [36, 39], we restrict our evaluation and analysis to the case $\epsilon = 0$ for the scope of this work.

## 6 Evaluation and Analysis

We begin by demonstrating that weakly hard real-time communication can serve networked control systems with meaningful guarantees throughout unstable network conditions. Thereafter, we provide a comparison with related state of the art and an in-depth analysis of our $(m, k)$-firm Elevation Policy in simulation environments with sporadic frame releases and 5G delay outliers.

[2]For Ethernet links, it captures the serialization delay of $f_i$ at port $(u,v)$, the propagation delay of the link $(u,v)$, and the processing delay of $f_i$ at bridge $v$ (i.e., $d_{min} \approx d_{max}$). For 5G links, it is determined by the latency guarantee with which $f_i$ is served by the 5G system (e.g., it may be derived from 5G delay histograms).

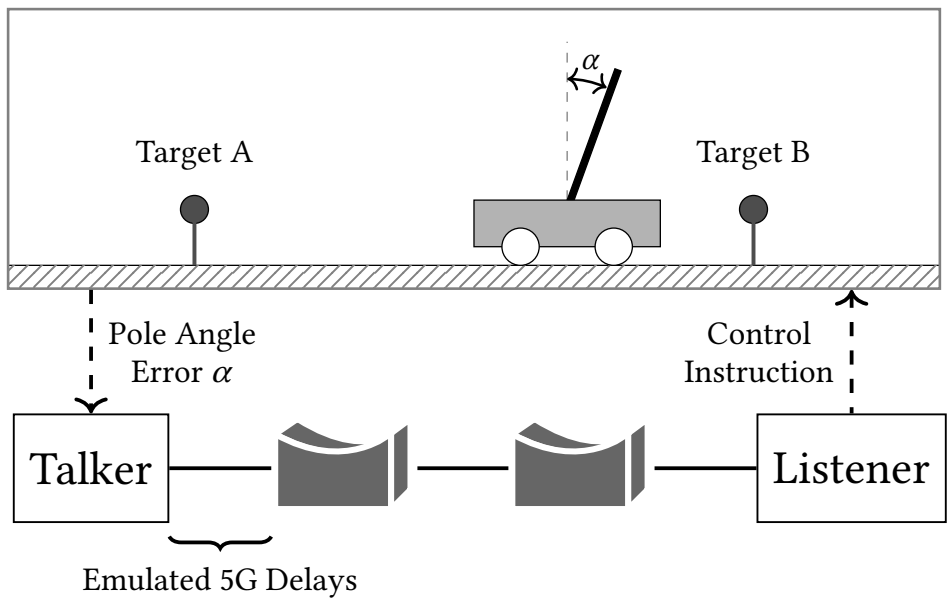

(a) Physical testbed setup

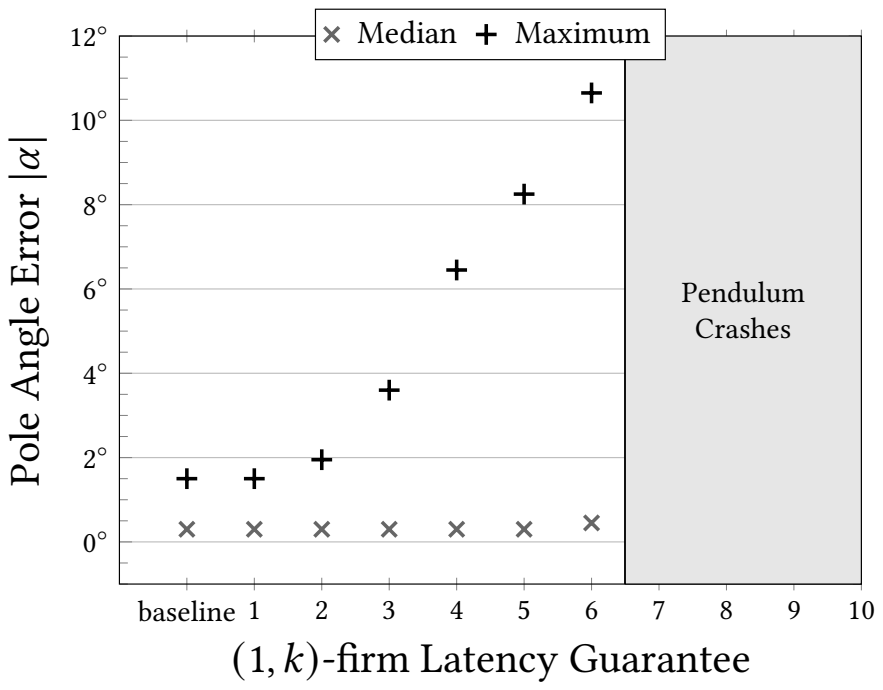

(b) Measured pole angle error

Fig. 6. Inverted pendulum (a) to evaluate the effects of $(m, k)$-firm latency guarantees on control systems (b).

## 6.1 Physical Testbed Benchmarks

We evaluate the effects of the $(m, k)$-firm Elevation Policy in a physical testbed where the control of an inverted pendulum is distributed via a TSN network (see Fig. 6a). The network is built using TSN-capable commercial-off-the-shelf (COTS) hardware: The talker and listener are two Linux machines, each equipped with a Intel I210 Ethernet controller to precisely schedule the frames' release times via the ETF Qdisc [34]. They are connected via two Kontron TSN bridges of the KSwitch D10 MMT series and use LinuxPTP [6] for time synchronization. On the first network link, the talker emulates 5G delays characteristics by delaying each frame according to the histograms measured by [25]. This allows us to switch between stable and unstable network conditions in a reproducible way that is not affected by external interference. In detail, we consider

i) *Stable epochs*, where the experienced 5G delays are upper-bounded by 7.71 ms. This corresponds to the 90 % percentile of the 5G uplink histogram from Figure 1b.
ii) *Unstable epochs* are initiated every 5 s, where this delay bound is violated for 10 consecutive frame transmissions. During this time, we sample 5G delays from the long-tail [7.71 ms, 13.89 ms]. Afterwards, the testbed switches back to stable conditions.

For reference, we also include *baseline* results where no unstable epochs occur.

The cart of the inverted pendulum switches between two alternating target positions (Target A and B) every 14 s to increase the difficulty of balancing the pendulum. The talker samples the pendulum's angle value $\alpha$ periodically every 75 ms and sends it to the listener. In turn, the listener computes the next control instruction to keep the pendulum upright, using the linear-quadratic regulator (LQR) from [28], and forwards it to the actuator of the pendulum. We evaluate the quality of control under different $(1, k)$-firm latency guarantees, for $k \in \{1, \ldots, 10\}$, by measuring the absolute pole angle error $|\alpha|$ of the inverted pendulum for each sampling period. For example, the $(1, 3)$-firm Elevation Policy is configured with the $\mu$-pattern of $\mu = 001$ such that the talker elevates the priority of every third frame during unstable epochs; otherwise, we use the default policy to discard any late frames (which is often required to shield other high-priority traffic). Each configuration is tested over 4 independent runs with a runtime of 15 min each, amounting to a total of 1 hour per configuration.

**Results:** Fig. 6b shows the median and maximum of the absolute pole angle error for each configuration. An error of $0^\circ$ means that the pendulum is in a perfectly upright position. The results provide two valuable insights: First, they clearly show that the median error is roughly the same for all configurations, whereas the maximum error shows an expected upward trend, as an increase in $k$ reflects the number of consecutive frame losses. This underlines the need to consider the worst-case behavior (i.e., the maximum error) for safety-critical system instead of their average-case performance. Second, the pendulum even starts to crash for configurations with $k > 6$. In particular, this is the case for TSN schedules that solely account for stable network conditions, e.g., as done by previous work [10]. Our $(m, k)$-firm Elevation Policy remedies this shortcoming and can be configured to serve meaningful QoS guarantees during unstable network conditions. For example, the observed system performance was always better than a maximal error of $2^\circ$ when the stream is served with $(1, 2)$-firm latency guarantees.

### 6.2 Evaluation in Wired TSN Scenarios with Sporadic Release Times

We continue with a more in-depth analysis of the $(m, k)$-firm Elevation Policy in a simulation environment with larger networks and more streams. In this section, we start with a special case where streams are solely affected by sporadic release times (e.g., due to unexpected compute delays). As there exist specialized scheduling techniques for wired TSN in this setting, we compare the performance of our $(m, k)$-firm Elevation Policy with that of E-TSN [44].[3] For clarity, the following uses *sporadic* and *isochronous* streams to denote streams with and without sporadic release times.

*6.2.1 Schedulability Analysis.* We start by analyzing the overall impact of the $(m, k)$-firm Elevation Policy on schedulability. The evaluation is based on a grid-structured network with $3\times4$ TSN bridges with 100 Mbps Ethernet links. Every TSN bridge connects to exactly one end-device. To quantify the effects on schedulability, we compute 100 stream sets per data point and count the number of feasible TSN schedules that are found. A feasible schedule has to satisfy the latency requirements of all isochronous streams while forwarding sporadic streams with minimal interference. We set a scheduling timeout of 30 min for E-TSN to find a feasible schedule. In contrast, our augmentation procedure of Section 5.2.2 operates as a single-shot procedure: It augments the primary schedule, which initially only accounts for isochronous streams, and verifies their latency bounds afterwards. We therefore set a scheduling timeout of 30 min for the primary schedule, whereas the runtime of our augmentation procedure is negligible in comparison (at most 60 ms in our benchmarks).

Each stream set consists of 24 isochronous streams and $N$ sporadic streams. For isochronous streams, we choose a random period of $F.period \in \{200\,\mu s, 400\,\mu s\}$ and a random latency requirement of $F.lat \in \{0.5F.period, 0.75F.period, F.period\}$. To constitute a fair comparison, sporadic streams are defined as in [44] based on a minimum inter-event time of either 200 µs or 400 µs; it is straight-forward to adjust our token bucket specification of Section 5.2.1 to this setting. Finally, both types of streams use a fixed frame size of 100 B with randomly selected talkers and listeners.

**Results:** Fig. 7 shows the schedulability in terms of the total number of feasible schedules found (total) and the portion of feasible stream sets that was rejected by the other approach (better). While the results show cases where our single-shot usage does not yield an optimal result, there remains a clear dominance of our $(m, k)$-firm Elevation Policy. This can be attributed to the overhead of the "prudent slot reservation" under E-TSN [44] to provide isochronous streams with additional transmission slots. In contrast, our augmentation of the primary schedule is fine-tuned to only impose the overhead needed for the exact token bucket specification of the sporadic streams.

[3] For completeness, we note that the authors of [44] do not provide a publicly available implementation. We therefore reimplemented their approach based on the scheduling constraints defined in their paper (see Appendix A).

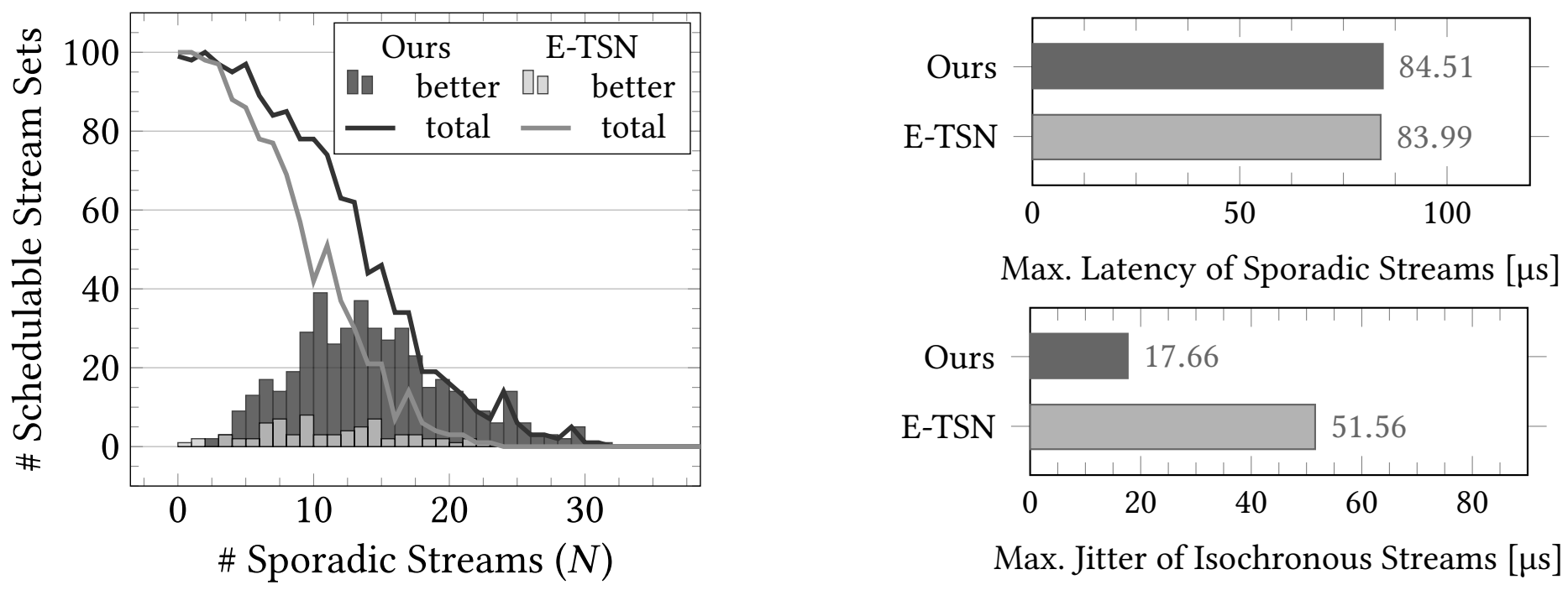

Fig. 7. Comparison of E-TSN [44] with the $(m, k)$-firm Elevation Policy (Ours) in terms of schedulability (left) and the worst-case impact on sporadic and isochronous streams (right).

*6.2.2 Worst-Case Analysis (Simulation).* Next, we consider a stream set with 24 isochronous and 24 sporadic streams that is schedulable under both E-TSN and the $(m, k)$-firm Elevation Policy. For validation and further analysis, we use OMNeT++ [41] with the INET extension [30] to simulate both configurations for a total simulation time of $2.5 \times 10^6$ hypercycles (i.e., 1000 s).

**Results.** Overall, the simulation validates that the latency requirements of all isochronous streams are always met; we provide the complete per-stream results in Appendix C. Fig. 7 shows two additional worst-case metrics that are of interest: First, the maximum latency of sporadic streams is nearly identical as sporadic streams are forwarded with an elevated priority under both configurations. Second, our augmentation procedure reduces the maximum jitter of the isochronous streams (i.e., the variation in the arrival time at the listener). It is indeed possible to show that the maximum jitter is upper-bounded by the prolongation step of Section 5.2.2. We therefore identify the reduced overhead of our schedule augmentation as the repeating cause for this improvement.

### 6.3 Evaluation in 5G-TSN Scenarios With 5G Delay Outliers

Next, we deploy the $(m, k)$-firm Elevation Policy in a simulated 5G-TSN network and compare it to the FIPS scheduler from [10]. The network is shown in Fig. 8a: It consists of two wired network partitions, i.e., the AGV-internal network (left) and the TSN backbone (right), which are internally equipped with 100 Mbps Ethernet links. The partitions are connected via a logical 5G-TSN bridge (using the DetCom [24] extension for OMNeT++) to allow the exchange of time-sensitive data. We configure FIPS to use the 99 % percentile of the 5G histograms from [25], i.e., $d_{max}$ is set to 9.98 ms and 11.037 ms for uplink and downlink streams. In turn, our $(m, k)$-firm Elevation Policy uses the computed schedule from FIPS as input and augments it as described in Section 5.2.

We consider a total of 100 streams, which we classify as wired traffic (20 streams) and wireless traffic (80 streams): Wired streams have both talker and listener located in the same wired network partition. They are characterized by a period of 5 ms and must be served with a worst-case network latency of 500 µs. These latency requirements are hard (i.e., with a $\mu$-pattern of $\mu = 1$) in that they have to be satisfied for every frame. In contrast, wireless streams traverse the logical 5G-TSN bridge either in uplink or downlink direction and have a more relaxed period and latency requirement of 20 ms. Their $(m, k)$-firm latency requirements are chosen randomly as follows: Half of the wireless streams specify an $(1, 3)$-firm requirement with a randomly chosen $\mu \in \{001, 010, 100\}$, while the other half has no real-time requirement during unstable network conditions (i.e., $\mu = 0$). The experiments are repeated for a total of $10^6$ hypercycles (approx. 5.5 h).

| Stream Type | PCP | Size | Period | Latency | $(m,k)$-firm Spec. |
|---|---|---|---|---|---|
| Wired | 6 | 100 B | 5 ms | 500 µs | $\mu = 1$ |
| Wireless | 5 | 100 B | 20 ms | 20 ms | $\mu \in \{0, 001, 010, 100\}$ |

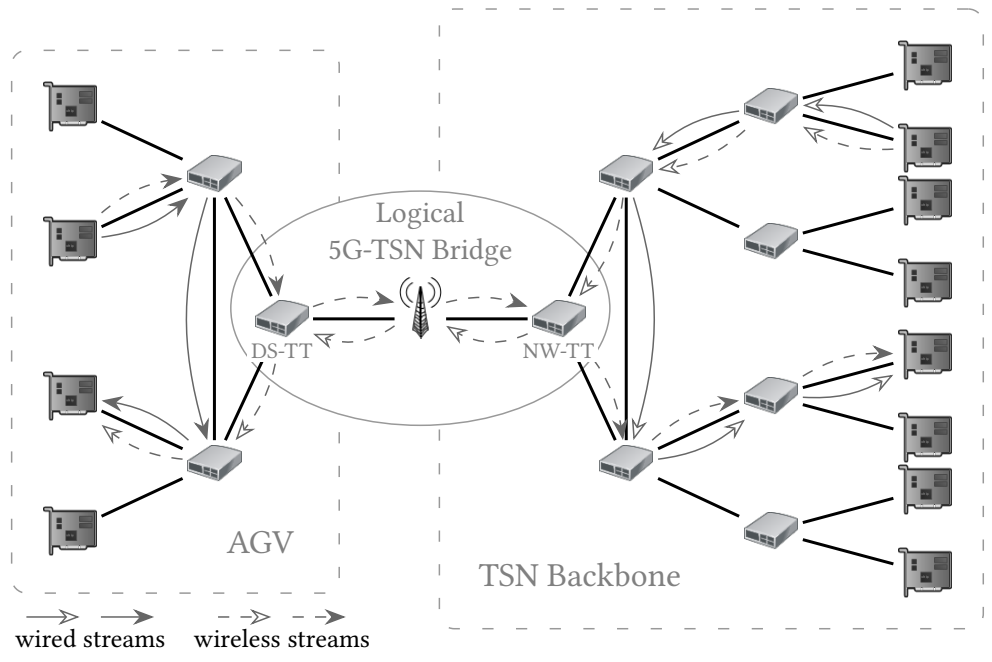

(a) Network topology and stream requirements

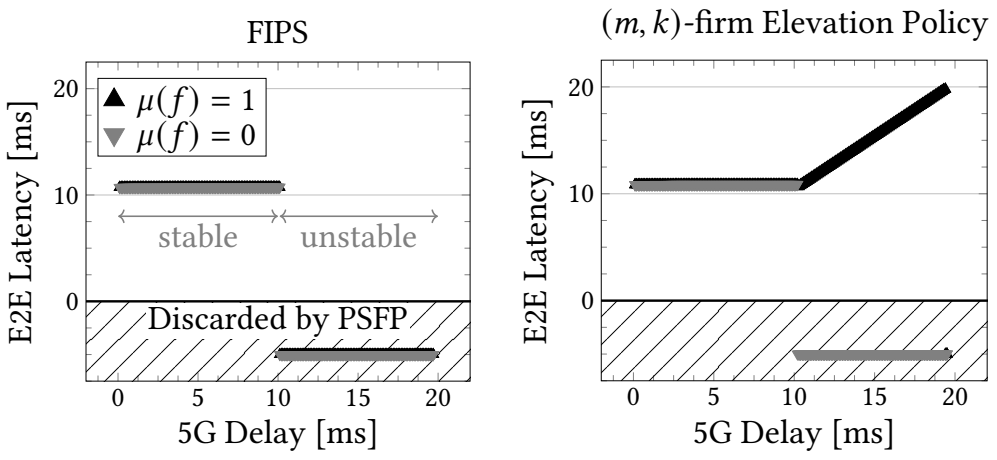

(b) Bounded 5G Channel Degradations (≤ 20 ms)

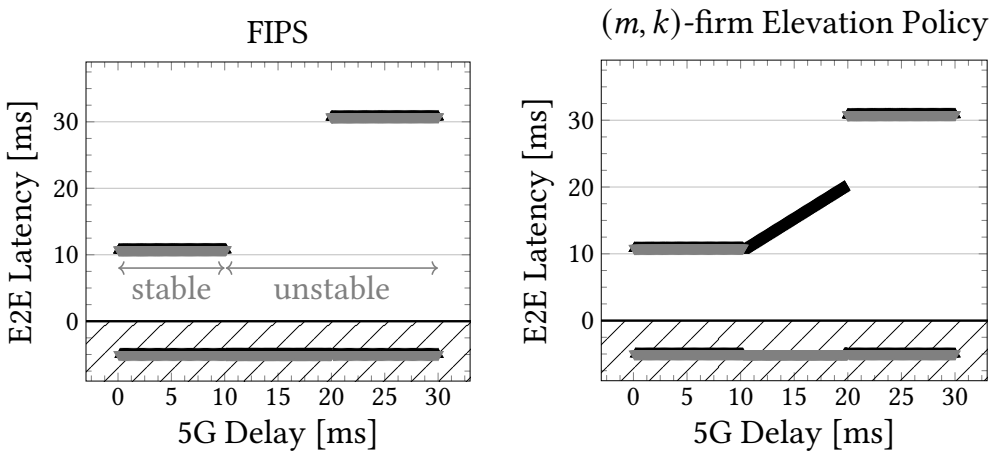

(c) Unbounded 5G Channel Degradations

Fig. 8. Simulation of a 5G-TSN network (a). We compare the real-time guarantees of FIPS [10] with the $(m,k)$-firm Elevation Policy under bounded (b) and unbounded (c) 5G channel degradations.

**Results:** Fig. 8b shows the end-to-end latency of a wireless (uplink) stream $F_1$ with an $(1,3)$-firm latency requirement under FIPS and the $(m,k)$-firm Elevation Policy. We use $\mu(f) = 1$ to denote frames that are eligible for priority elevation under the configured $\mu$-pattern. First and foremost, the results show that FIPS does not provide any guarantees for any frame $f \in F_1$ with a 5G delay above the configured 99 % percentile (i.e., 9.98 ms). Instead, it strictly discards $f$ via PSFP to shield the real-time guarantees of other traffic [10], e.g., to shield the wired streams with hard real-time requirements. This policy can significantly impair the application performance, as previously demonstrated in Section 6.1. In contrast, our $(m,k)$-firm Elevation Policy accounts for the configured $\mu$-pattern and elevates the priority of late frames $f$ if and only if $\mu(f) = 1$. Thereby, it can endure 5G delays close to the stream's latency requirement of 20 ms. At the same time, the necessary augmentation of the schedule from FIPS results in a small latency increase of below 190 µs during stable network conditions (i.e., when the 5G delays lie within the 99 % percentile).

Lastly, we discuss the effects when the 5G channel degradations can even exceed the 20 ms period of the stream. Fig. 8c shows a second uplink stream $F_2$, which is scheduled under the same TSN schedules as before but can experience 5G delays of up to 30 ms. The results show, for both configurations, that there can be unintended PSFP-induced frame loss for a frame $f^{(i)} \in F_2$ with 5G delays in [0 ms, 10 ms]. This is because the IEEE 802.1Q tag does not contain sequence numbers, whereby PSFP at the NW-TT cannot differentiate between $f^{(i)}$ and $f^{(i-1)} \in F_2$ from the previous period with 5G delays in [20 ms, 30 ms]. Still, we note that there remains a meaningful fault-isolation between streams as these faults stay isolated to $F_2$ and do not spread to other streams.

## 7 Discussion

Before concluding this paper, we discuss limitations and practical aspects that should be considered before deploying the $(m,k)$-firm Elevation Policy and provide directions for future work.

*Limitations Due to 5G Modeling Assumptions.* We recognize that modeling the 5G system as abstractly as a logical 5G-TSN bridge has both advantages and disadvantages. On the one hand, we argue that the intolerance of time-driven schedules against delay outliers is caused by their reliance on idealistic 5G delay models (see Section 4.2). The $(m, k)$-firm Elevation Policy avoids such dependencies to realize a fallback strategy that increases the robustness against unexpected delay outliers. Moreover, our evaluations demonstrated that this makes the $(m, k)$-firm Elevation Policy applicable to broader TSN scenarios where frames experience sporadic release times (e.g., due to unexpected compute delays or time synchronization errors). On the other hand, we perceive that there remain vastly unexplored co-design options that may provide better performance or stronger real-time guarantees by coordinating the fallback policy across all domains (including the application domain). As co-design in general remains an open key issue to achieve true end-to-end dependability, we plan to further explore this direction in our future work.

*Configuration Options for the $(m, k)$-firm Elevation Policy.* Each stream can specify its weakly hard real-time requirements via a $\mu$-pattern. While this work assumed that the $\mu$-patterns are predetermined, common survival time requirements of applications provide more flexibility. For instance, a requirement that one out of every three consecutive frames meets its requirement [31] solely constrains $\mu \in \{001, 010, 100\}$ and allows the scheduler to chose. To best utilize our $(m, k)$-firm Elevation Policy, the $\mu$-patterns should be selected to evenly spread-out elevated traffic. For instance, for 6 streams with the same period (say 20 ms) and the above survival time requirement, choosing 6 times the same $\mu$-pattern of $\mu = 100$ will result in an unnecessary resource allocation during the interval $[20\,\text{ms}, 60\,\text{ms}]$ where no elevated traffic would ever arrive. Our token bucket model for elevated traffic instead assumes an evenly spread-out arrival (see Section 5.2.1), which is better matched by a uniform selection across $\mu \in \{001, 010, 100\}$.

*Deploying the $(m, k)$-firm Elevation Policy at Scale.* In general, any provisioning of fine-grained real-time guarantees encounters fundamental challenges when scaling up the number of streams. We argue that the required PSFP state information — which specifies when to forward, elevate or discard frames at each TSN bridge — is the most stringent factor to limit scalability. Practical fault tolerance engineering can mitigate this limitation to a certain degree: First, multiple streams can be aggregated into a service class (e.g., similar to DiffServ [43]) to reduce the PSFP state per TSN bridge. Second, the deployment of PSFP can be reduced exclusively to TSN bridges that follow network segments prone to delay outliers (e.g., as in Fig. 3a). However, it is clear that both relaxations come at a certain price, i.e., more coarse-grained QoS guarantees in the first case and the creation of common failure domains due to unprotected links in the second case. For the scope of this work, we focused on applications that need per-stream guarantees with the highest degree of fault tolerance.

## 8 Conclusion

This work addresses the problem that modern 5G-TSN networks remain prone to abrupt delay outliers. The underlying causes are often related to external factors, e.g., 5G line-of-sight blockage, which makes them difficult to predict and do not leave enough time to reactively reconfigure the network. We introduce the $(m, k)$-firm Elevation Policy to overcome this challenge. It acts as a fallback mechanism to the primary time-driven schedule and elevates the priority of late frames if they are at risk of violating the application's survival time requirements. Our evaluations showed that these guarantees are essential for networked control systems to uphold quality of control during unstable network conditions. Moreover, we demonstrated that the $(m, k)$-firm Elevation Policy stays complementary to existing TSN schedulers, while imposing only a small resource overhead during stable network conditions. Thereby, this work provides a robust but light-weight fallback mechanism that achieves meaningful real-time guarantees during unstable network conditions.

*Future Work.* In addition to the future research directions highlighted in Section 7, we also plan to explore similar fallback strategies for traffic types other than periodic traffic.

*Ethics.* No ethical concerns were raised for this work.

## Acknowledgment

This work was supported by the European Commission through the H2020 project DETERMINISTIC6G (Grant Agreement no. 101096504). We thank Melanie Heck and the DETERMINISTIC6G consortium for their valuable feedback.

## A Artifact Summary

Several prototypes were created for this paper and will be made publicly available on GitHub. The corresponding datasets for our evaluations will be made available on Zenodo. We plan to publish the results of all intermediate steps (e.g. our generated topologies and stream sets, our scheduler results as well as our simulation results) together with scripts to extract the relevant data used in our evaluation. We provide a brief summary in the following:

*Proof of Concept (PoC) Implementation.* Our artifacts include the implementation of the $(m, k)$-firm Elevation Policy in OMNeT++ and the schedule augmentation procedures.

*Physical Testbed.* Our published artifacts will consists of the implementation of talker and listener applications, as well as the TSN bridge configuration, our measurement data, and evaluation scripts.

*Wired TSN Scenarios with Sporadic Release Times.* The evaluation is run on an automated pipeline and will also be published in its entirety. First, this includes a scenario generator to generate network topologies and stream sets. Second, a conventional constraint-programming-based TSN scheduler (based on [7]) is used to synthesize the primary schedule. Third, we provide scripts to execute our PoC augmentation procedure, or alternatively, our E-TSN implementation (based on the prudent reservation approach from [44]). Lastly, our artifacts include the code to generate, execute and analyze OMNeT++ simulation scenarios based on the scheduler results.

*5G-TSN Scenarios With 5G Delay Outliers.* The evaluation runs on an analogous pipeline as above. The only difference is that the primary schedule is generated using the FIPS scheduler from [10].

## B Generalized Schedule Augmentation in Multi-Hop Settings

The main difference to Section 5.2.2 is that the general traversal of the scheduled transmissions requires a slightly more complicated data structure that captures all temporal scheduling dependencies. For this purpose, we introduce so-called *Transmission Graphs.*

### B.1 Transmission Graphs

Inspired by the Disjunctive Graph Model — which is well-known in the job-shop scheduling literature and was first introduced by [33] — we define Transmission Graphs to encode the initial TSN schedule. The following uses Fig. 9 to illustrate the construction for our example from Fig. 4.

For a given input schedule $C$, we define its transmission graph as a weighted directed graph $\mathcal{G}_C = (\mathcal{V}, \mathcal{E}, w)$. The vertices $\mathcal{V}$ consist of

$$\mathcal{V} = \left\{O_f^i \mid \exists F \in \mathcal{F} : f \in F \text{ and } (v_F^i, v_F^{i+1}) \in F.route\right\} \cup \{\emptyset, *\}.$$

Vertices of the form $O_f^i$ represent the transmission time of the frame $f$ at the $i$th egress port in $F.route$ (i.e., $1 \leq i < n_F$) and are called *operations.* Moreover, $\emptyset$ and $*$ are specially denoted *source* and *sink* vertices and we often write $O_f^0 = \emptyset$ and $O_f^{n_F} = *$ for convenience.

In turn, the edges $\mathcal{E}$ are commonly categorized into conjunctive and disjunctive edges [33]: A *conjunctive edge* $C_f^i = (O_f^i, O_f^{i+1})$ is added for each frame $f$ and each $0 \leq i < n_F$, with weight

$$w(C_f^i) = \begin{cases} f.release, & \text{for } i = 0, \\ d_{max}((v_f^i, v_f^{i+1}), f), & \text{for } 0 < i < n_F. \end{cases}$$

If $i = 0$, the weight equals the time when $f$ is released by the application, e.g., shown in Fig. 9 for the outgoing edges from $\emptyset$. If $i > 0$, the weight equals the delay between the transmission start of $f$ at $v_f^i$ and the time $f$ is enqueued at $v_f^{i+1}$ (as in Section 5.2.2). For example, Fig. 9 shows this for

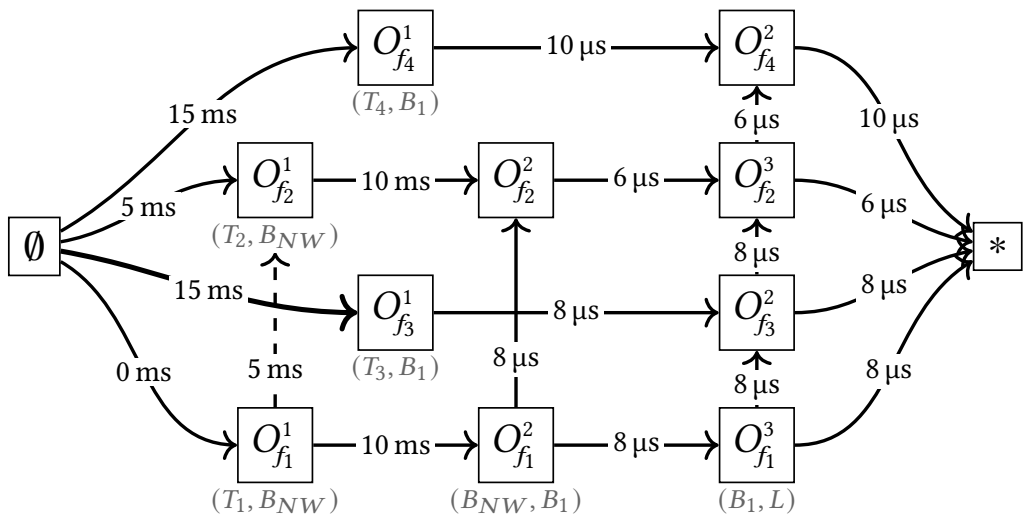

| $F$ | $F.size$ | $F.phase$ | $F.pcp$ | $[d_{min}(F), d_{max}(F)]$ Eth* | 5G |
|---|---|---|---|---|---|
| $F_1$ | 800 bit | 0 | 4 | 8 µs | [0 ms, 10 ms] |
| $F_2$ | 600 bit | 5 ms | 4 | 6 µs | [5 ms, 10 ms] |
| $F_3$ | 800 bit | 15 ms | 5 | 8 µs | – |
| $F_4$ | 1000 bit | 15 ms | 6 | 10 µs | – |

* assuming 100 Mbps Ethernet links with $d_{min} = d_{max}$

Fig. 9. Transmission graph for the initial $\mathcal{S}_{GCL}$ from Fig. 4 (left) under given stream parameters (right). For simplicity, we only show one frame per stream and assume zero propagation and processing delay.

- *Ethernet Links:* The conjunctive edge $C^2_{f_2} = (O^2_{f_2}, O^3_{f_2})$ accounts for the serialization delay 600 bit/100 Mbps = 6 µs of $f_2$ at the port $(B_{NW}, B_1)$.
- *5G Links:* The conjunctive edge $C^1_{f_1} = (O^1_{f_1}, O^2_{f_1})$ accounts for the 5G delay budget [0 ms, 10 ms] to exceed the 99 % quantile of the 5G uplink delay histogram of Fig. 1b.

The conjunctive edges thus ensure that $f$ is only scheduled for transmission at the port $(v^{i+1}_f, v^{i+2}_f)$ after being received, processed, and enqueued at $v^{i+1}_f$.

A *disjunctive edge* $D^{i,j}_{f \to f'} = (O^i_f, O^j_{f'})$ is added for each pair of frames $f \in F, f' \in F'$ (with $f \neq f'$) that share a common link, i.e., there exists $i$ and $j$ with $(v^i_f, v^{i+1}_f) = (v^j_{f'}, v^{j+1}_{f'})$. The orientation $f \longrightarrow f'$ is selected based on their ordering in $C$, i.e., $f \longrightarrow f'$ is chosen if and only if $C$ schedules the transmission of $f$ at the common link to start earlier than that of $f'$. By choosing the weight

$$w(D^{i,j}_{f \to f'}) = d_{max}((v^i_f, v^{i+1}_f), f),$$

we ensure that the transmission of $f'$ is scheduled after $f$ completed its transmission. In Fig. 9, we only draw $D^{i,j}_{f \to f'}$ if $f'$ is directly scheduled after $f$, e.g., the sequence $(O^3_{f_1}, O^2_{f_3}, O^3_{f_2}, O^2_{f_4})$ captures all six disjunctive edges at the port $(B_1, L)$. This allows for an efficient implementation that only needs to stores the index of $O^i_f$ in these sequences instead of all disjunctive edges.

It remains to encode the FIFO queuing policy when $f$ and $f'$ share the same queues. To this end, we proceed as follows for every disjunctive edge $D^{i,j}_{f \to f'} \in \mathcal{E}$ with $f.pcp = f'.pcp$:

- If $f$ and $f'$ share the previous link (i.e., $v^{i-1}_f = v^{j-1}_{f'}$), we verify that $D^{i-1,j-1}_{f \to f'}$ is contained in $\mathcal{E}$. This is because the inverse orientation $f' \longrightarrow f$ would yield a FIFO violation by requiring $f$ to overtake $f'$ in the same egress queue at $(v^i_f, v^{i+1}_f)$. In Fig. 9, this is required for the orientation $f_1 \longrightarrow f_2$ along the consecutive hops $(B_{NW}, B_1)$ and $(B_1, L)$.
- Otherwise, we add the *FIFO edge* $D^{i-1,j-1}_{f \to f'}$ with weight

$$w(D^{i-1,j-1}_{f \to f'}) = d_{max}((v^{i-1}_f, v^i_f), f) - d_{min}((v^{j-1}_{f'}, v^j_{f'}), f'),$$

where $d_{min} \in [0, d_{max}]$ denotes a known lower bound of the transmission delay. This defers the operation $O^{j-1}_{f'}$ to ensure $f$ is always the first to be enqueued at $(v^i_f, v^{i+1}_f)$. In Fig. 9, this is shown for the dashed edge $(O^1_{f_1}, O^1_{f_2})$ to ensure the correct arrival order of $f_1$ and $f_2$ at $B_{NW}$.

Finally, we note that Transmission Graphs are very versatile and can easily be extended to encode additional constraints, e.g., to account for fault-isolation constraints [7] and frame batching [10].

## B.2 Augmenting TSN Schedules

Let $\mathcal{G}_C$ be the Transmission Graph of the initial TSN schedule $C$. We consider a topological sort $\sigma : \mathcal{V} \longrightarrow \{1, \dots, |\mathcal{V}|\}$, i.e., $\sigma(O^i_f) < \sigma(O^j_{f'})$ for $(O^i_f, O^j_{f'}) \in \mathcal{E}$, which can be computed via a simple

DFS traversal of $\mathcal{G}_C$. This allows for an iterative approach, where the $n$th iteration adds GCL and PSFP entries for the operation $O_f^i \in \mathcal{V}$ with $\sigma(O_f^i) = n$. We maintain two global variables:

- the critical cost $C(O_f^i)$ that denotes the cost of the longest path from $\emptyset$ to $O_f^i$, and
- the prolongation delay $\theta(u, v)$ that captures the induced delay due to elevated traffic.

Initially, all variables are set to zero. The $n$th iteration then consists of the following steps:

*1) Update Prolongation Delay:* We first note that the topological sort ensures that $O_f^i$ can only be influenced by preceding operations (i.e., by $O_{f'}^j$ with $\sigma(O_{f'}^j) < n$). Therefore, the critical cost $C(O_f^i)$ will not be modified by later iterations and can safely be used as the earliest transmission start of $O_f^i$. In contrast, the latest transmission start is influenced by the maximum burst size of elevated traffic at $(v_f^i, v_f^{i+1})$. We update $\theta$ as follows:

$$\theta(v_f^i, v_f^{i+1}) \leftarrow \max\left\{\theta(v_f^i, v_f^{i+1}),\ C(O_f^i) + \frac{b(v_f^i, v_f^{i+1})}{(v_f^i, v_f^{i+1}).bitrate - r(v_f^i, v_f^{i+1})}\right\}.$$

*2) Add GCL and PSFP Entries:* An GCL entry is added to $\tilde{\mathcal{S}}_{GCL}(v_f^i, v_f^{i+1})$ that opens the gate of the traffic class $f.pcp$ during the interval

$$\left[C(O_f^i),\ \theta(v_f^i, v_f^{i+1}) + d_{max}((v_f^i, v_f^{i+1}), f)\right).$$

Analogously, an PSFP entry is added to $\mathcal{R}(v_f^{i+1})$ that *forwards* $f$ if it arrives during

$$\left[C(O_f^i) + d_{min}((v_f^i, v_f^{i+1}), f),\ \theta(v_f^i, v_f^{i+1}) + d_{max}((v_f^i, v_f^{i+1}), f)\right).$$

Let $F$ denote the corresponding stream of $f$. In case the $\mu$-pattern of $F$ specifies that $f$ can be elevated under the $(m, k)$-firm Elevation Policy, we also add an PSFP entry to $\mathcal{R}(v_f^{i+1})$ that *elevates* $f$ if it arrives during

$$\left[\theta(v_f^i, v_f^{i+1}) + d_{max}((v_f^i, v_f^{i+1}), f),\ f.release + f.lat\right).$$

In all other cases, the GCL and PSFP gates remain closed.

*3) Update Critical Cost (Deferment):* Finally, we update the variables of the succeeding operations $O_{f'}^j$ (i.e., with $(O_f^i, O_{f'}^j) \in \mathcal{E}$) depending on their relation to $O_f^i$: For disjunctive edges $D_{f \to f'}^{i,j}$ with $f'.pcp \leq f.pcp$, we allow an overlap in the transmission slots of $O_f^i$ and $O_f^j$ and update

$$C(O_{f'}^j) \leftarrow \max\left\{C(O_{f'}^j),\ C(O_f^i) + w(D_{f \to f'}^{i,j})\right\}$$

$$\theta(v_f^i, v_f^{i+1}) \leftarrow \theta(v_f^i, v_f^{i+1}) + \frac{d_{max}((v_f^i, v_f^{i+1}), f) \times r(v_f^i, v_f^{i+1})}{(v_f^i, v_f^{i+1}).bitrate - r(v_f^i, v_f^{i+1})}$$

For all other outgoing edges — i.e., for conjunctive edges, disjunctive edges with $f'.pcp > f.pcp$, and FIFO edges — the operation $O_{f'}^j$ is deferred via

$$C(O_{f'}^j) \leftarrow \max\left\{C(O_{f'}^j),\ \theta(v_f^i, v_f^{i+1}) + w(D_{f \to f'}^{i,j})\right\}.$$

## C Per-Stream Simulation Results

The simulation results are shown in Fig. 10, validating that the latency requirements of all isochronous streams can be guaranteed by both E-TSN and our $(m, k)$-firm Elevation Policy. For a quantitative comparison between both approaches, we refer the reader to Section 6.2.

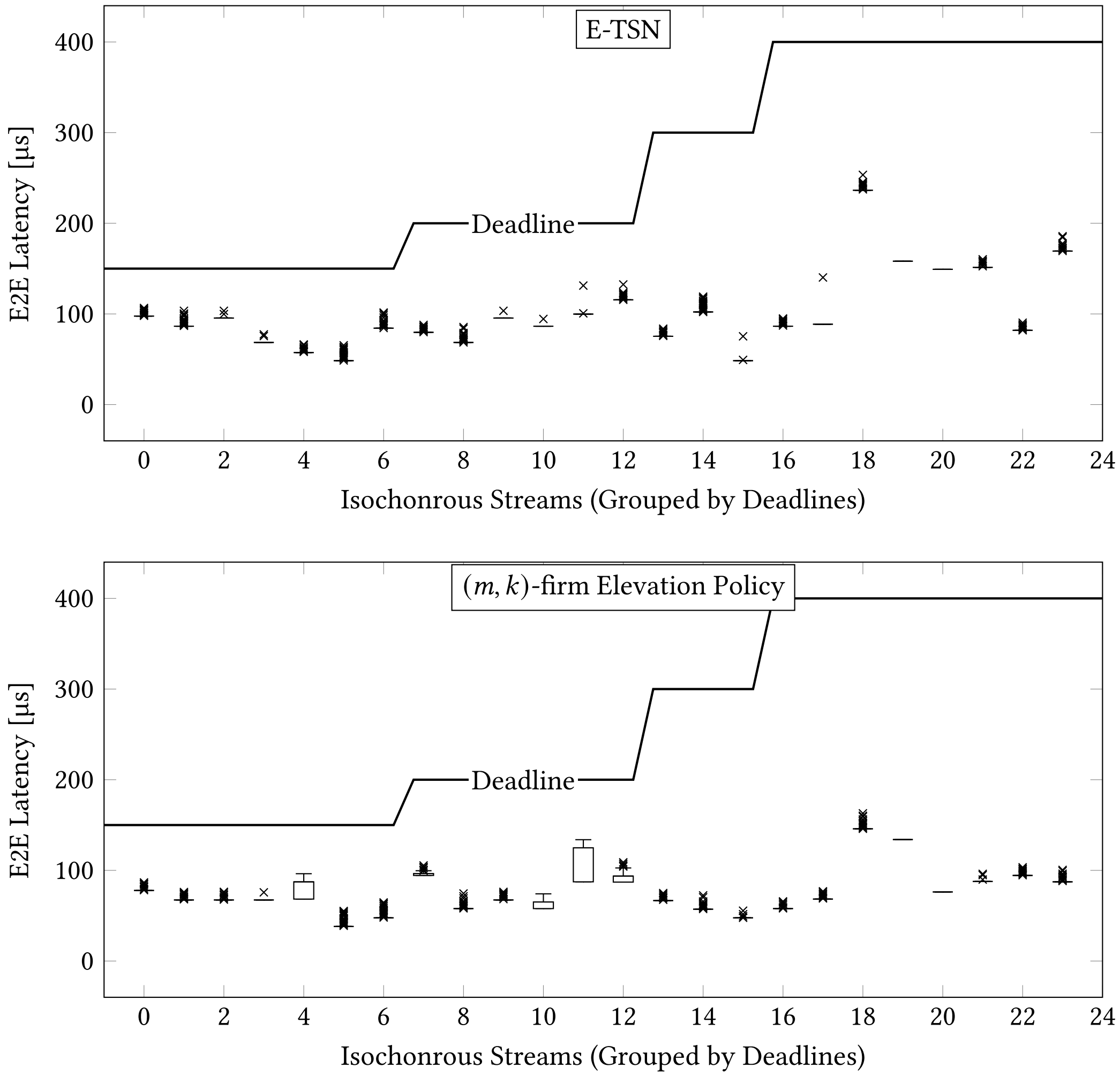

Fig. 10. Per-Stream Simulation Results of E-TSN and our $(m, k)$-firm Elevation Policy.